\documentclass{emulateapj}
\usepackage{graphicx}
\usepackage{color}
\usepackage{amssymb}

\newcommand{\kms}{\,{\rm km\,s^{-1}}}
\newcommand{\ms}{\,{\rm m\,s^{-1}}}
\newcommand{\kmsmpc}{\,{\rm km\,s^{-1}\,Mpc^{-1}}}
\newcommand{\msun}{\,{\rm M_\odot}}
\newcommand{\rvir}{r_{\rm vir}}
\newcommand{\mvir}{M_{\rm halo}}
\newcommand{\msub}{M_{\rm subh}}
\newcommand{\vcmax}{v_{\rm c,max}}
\newcommand{\rvcmax}{r_{\rm \vcmax}}
\newcommand{\vcmaxhost}{v_{\rm c,max,host}}
\newcommand{\vcmaxsub}{v_{\rm c,max,sub}}
\newcommand{\rlagrange}{R_{\mathcal{L},\rm vir}}
\newcommand{\f}{\frac}

\newcommand\be{\begin{equation}}
\newcommand\ee{\end{equation}}

\lefthead{Diemand, Kuhlen, \& Madau}
\righthead{Early SUSY-CDM substructure}

\begin{document}
\submitted{ApJ in press}

\title{Early supersymmetric cold dark matter substructure}
\author{J\"urg Diemand, Michael Kuhlen, \& Piero Madau}
\affil{Department of Astronomy \& Astrophysics, University of California, Santa Cruz, CA 95064}
\email{diemand@ucolick.org.}

\begin{abstract}
Earth-mass ``microhalos'' may be the first
objects to virialize in the early universe.
Their ability to survive the hierarchical
clustering process as substructure in the
larger halos that form subsequently has
implications for dark matter detection
experiments. We present a large N-body simulation of 
early substructure in a supersymmetric cold
dark matter (SUSY-CDM) scenario
characterized by an exponential cutoff
in the power spectrum at $M_c=10^{-6}\,\msun$.
The simulation resolves a $0.014\,\msun$
parent ``SUSY'' halo at $z=75$ with 14
million particles. On these scales the
effective index of the power spectrum
approaches $-3$, and a range of mass
scales collapse almost simultaneously.
Compared to a $z=0$ galaxy cluster 
substructure within our SUSY host is less
evident both in phase-space and in physical
space, and it is less resistant against 
tidal disruption. As the universe expands
by a factor of 1.3, we find that between 20
and 40 percent of well-resolved SUSY 
substructure is destroyed, compared to
only $\sim 1$ percent in the low-redshift
cluster. Nevertheless SUSY substructure
is just as abundant as in $z=0$ galaxy
clusters, i.e. the normalized mass and
circular velocity functions are very 
similar. The dark matter self-annihilation
$\gamma$-ray luminosity from bound subhalos
and other deviations from a smooth spherical
configuration is at least comparable to the
spherically-averaged signal in the SUSY host,
and at least three times larger than the
spherically-averaged signal in the cluster host.
Such components must be taken into account when
estimating the total cosmological extragalactic
$\gamma$-ray annihilation background. The relative
contribution of bound substructure alone to the total
annihilation luminosity is about four times smaller
in the SUSY host than in the $z=0$ cluster because of
the smaller density contrast of sub-microhalos.
\end{abstract}


\section{Introduction}

The key idea of the standard cosmological paradigm for the formation
of structure in the universe, that primordial density fluctuations
grow by gravitational instability driven by cold, collisionless dark
matter (CDM), is constantly being elaborated upon and explored in
detail through supercomputer simulations, and tested against a variety
of astrophysical observations. The leading candidate for dark matter
(DM) is the neutralino, a weakly interacting massive particle
predicted by the supersymmetry (SUSY) theory of particle
physics. While in a SUSY-CDM scenario the mass of bound DM systems
(halos) may span about twenty order of magnitudes, from the most
massive galaxy clusters down to Earth-mass clumps
\citep*[e.g.][]{Green2004}, it is the smallest DM microhalos that
collapse first, and only these smallest scales are affected by the
nature of the relic DM candidate.

Recent numerical simulations of the collapse of the earliest and
smallest gravitationally bound CDM clumps
\citep*{Diemand2005susy,Gao2005} have shown that tiny virialized
microhalos form at redshifts above 50 with internal density profiles
that are quite similar to those of present-day galaxy clusters. At
these epochs a significant fraction of neutralino DM has already been
assembled into non-linear Earth-mass overdensities. If this first
generation of dark objects were to survive gravitational disruption
during the early hierarchical merger and accretion process -- as well
as late tidal disruption from stellar encounters
\citep*{Zhao2005,Berezinsky2005} -- then over $10^{15}$ such clumps
may populate the halo of the Milky Way \citep{Diemand2005susy}. The
nearest microhalos may be among the brightest sources of $\gamma$-rays
from neutralino annihilation. As the annihilation rate increases
quadratically with the DM density, small-scale clumpiness may enhance
the total $\gamma$-ray flux from nearby extragalactic systems (like
M31), making them detectable by the forthcoming {\it GLAST} satellite
or the next-generation of air Cerenkov telescopes
\citep*[e.g.][]{Oda2005}. Dark matter caustics might occur
within the first generation of halos. Caustics are not resolved in N-body simulations
like the ones discussed here and might have significant effects on
DM detection experiments \citep[e.g.][]{Mohayaee2005}.

The possibility of observing the fingerprints of the smallest-scale
structure of CDM in direct and indirect DM searches hinges on
the ability of microhalos to survive the hierarchical clustering
process as substructure within the larger halos that form at later
times. In recent years high-resolution N-body simulations have enabled
the study of gravitationally-bound subhalos with $\msub/\mvir
\gtrsim 10^{-6}$ on galaxy (and galaxy cluster) scales
\citep*{Moore1999,Klypin1999,Stoehr2003,Diemand2004sub,Gao2004,Reed2005}. The main
differences between these subhalos -- the surviving cores of objects
which fell together during the hierarchical assembly of galaxy-size
systems -- and the tiny sub-microhalos mentioned above is that on the
smallest CDM scale the effective index of the linear power spectrum of
mass density fluctuations is close to $-3$. In this regime typical
halo formation times depend only weakly on halo mass, the capture of
small clumps by larger ones is very rapid, and sub-microhalos may be
more easily disrupted. Yet previous simulations that have resolved
subhalos above $1000\,\msun$ down to $z=4$ \citep{Moore2001} and above
$10\,\msun$ down to $z=49$ \citep{Gao2005} have found that the
substructure abundance in early DM halos is quite similar to that of their
present-day counterparts.

In this paper we extend this range further down towards the
free-streaming cutoff mass of $10^{-6}\,\msun$ by using a mass
resolution of $10^{-9}\msun$ in a large high-resolution $N$-body
simulation of early substructure in SUSY-CDM.  The simulation
resolves a $0.014\,\msun$ host halo at $z=75$ with 14 million particles
within its virial radius. The mass resolution and power spectrum are the same
as in one of the runs presented in \citet{Diemand2005susy}, which
focused on a region of average density. The new simulation discussed
in this work has a high resolution region that covers an eight times
larger initial volume [120 (comoving pc)]$^3$ and is centered on the
highest-$\sigma$ peak of the same 3 comoving kpc parent simulation
cube.

The plan of the paper is as follows. In \S~2 we describe the initial conditions, 
numerical methods, and parameters of our simulations. In \S~3 
we present the properties of our SUSY parent halo and
its substructure, comparing them to a low-redshift massive galaxy cluster
simulated at comparable resolution, and discuss the relative
abundance and survival probabilities of substructure in both
simulations. In \S~4 we estimate the dark
matter self-annihilation $\gamma$-rays signal from substructure at $z=75$ and 
$z=0$. Finally, we summarize our conclusions in \S~5.
Details about finding early CDM substructures both with SKID and a 
new phase-space friends-of-friends (6DFOF) algorithm are given in the
Appendix.

\section{N-body simulations}
\label{sec:simulations}

The results of this paper are based on a comparison between two dark
matter simulations on vastly different scales. One simulation
(``SUSY'') resolves DM substructure all the way down to the
free-streaming scale for a generic supersymmetric particle of mass
$m_X=100$ GeV, and is designed to model the formation of the
smallest and earliest dark microhalos in the universe. The other
simulation (``B9'') models the formation of a $5.9\times 10^{14}\,\msun$
cluster at the present epoch, and was previously discussed in
\citet*{Diemand2004cluster}.

Both simulations were performed with PKDGRAV
\citep{Stadel2001,Wadsley2004} and employed the multiple mass particle
grid initial conditions as described in \citet{Diemand2004cluster}. The
SUSY run has a comoving box size of 3 kpc and was initialized at
$z=456$ using the GRAFICS package \citep{Bertschinger2001} with a DM
power spectrum $P(k)$ for a 100 GeV neutralino (kinetic decoupling
occurs at $T_d=28$ MeV) given in \citet{Green2004}\footnote{As in
\citet{Diemand2005susy} we use model ``A'' of the first
preprint version of \citet{Green2004} (astro-ph/0309621v1).}. This
power spectrum is close to $P(k) \sim k^{-3}$ and exhibits a free
streaming cutoff at $k = 1.65$ (comoving pc)$^{-1}$, corresponding to
a mass scale of $1.1 \times 10^{-6} \msun$. The initial coupling
between DM and radiation produces acoustic oscillations 
that leads to a slightly higher cutoff mass of $4.5 \times 10^{-6} \msun$
\citep[see eq. 26 in][]{Loeb2005}, but we have neglected this
correction in the present work.

The SUSY simulation is centered on a rare high-$\sigma$ peak, which is
covered by a 120 comoving pc high-resolution region consisting of 64
million DM particles. This results in a mass per particle of $9.8
\times 10^{-10} \msun$, which ensures that even objects at the
free-streaming limit are resolved with
$\sim 1000$ particles. At $z=75$ the most massive halo in the box has 
a virial mass\footnote{The ``virial'' mass is defined to enclose
a mean density 200 times the mean background density in the high redshift SUSY run.
For the $z=0$ cluster, it is 368 times the mean background density.}
of $\mvir=0.014\,\msun$, which corresponds to a 3.5-$\sigma$
fluctuation. We have adopted the standard concordance $\Lambda$CDM
cosmology with parameters $\Omega_M = 0.268$, $\Omega_\Lambda = 0.732$, $H_0
= 71\,\kmsmpc$, $n=1$, and a power spectrum normalized to 
$\sigma_8\equiv\sigma(z=0,r=8\,h^{-1}\,\rm Mpc)=0.9$ in the SUSY and $0.7$ 
in the B9 run. The parameters of both simulations are summarized 
in Table \ref{tab:parameters}.

\begin{table}
\begin{center}
\caption{Simulation parameters}
\label{tab:parameters}
\begin{tabular}{ccccc}
\hline
\hline
$z_i$ & $z_f$ & $\epsilon$     & $N_{\rm hires}$ & $m_{\rm p,hires}$ \\
      &       & (comoving units) &               & ($\msun$) \\
\hline
455.8 & 75.1 & 0.00216 pc & $6.4 \times 10^7$ & $0.98 \times 10^{-9}$ \\
40.3 & 0.0 & 4.8 kpc & $2.8 \times 10^7$ & $5.2\times 10^{7}$ \\
\end{tabular}
\end{center}
\tablecomments{Initial and final redshifts $z_i$ and $z_f$, (spline)
softening length $\epsilon$, and number $N_{\rm hires}$ and masses
$m_{\rm p,hires}$ of high resolution particles, for the SUSY ({\it top}) 
and B9 ({\it bottom}) runs.}
\end{table}

\section{Halo substructure}
\label{sec:Halo}

The $k^{-3}$ dependence of the power spectrum near the SUSY cutoff
translates into a $\sigma(M)$ (the linear theory rms density 
fluctuations extrapolated to the present-day in a sphere containing a 
a mass $M$), that is nearly independent of mass.
This implies that structures are simultaneously collapsing over a wide
range of masses and has important consequences for the nature of DM
(sub-) structure near the SUSY cutoff, as we show below.

\subsection{Structure formation on SUSY scales} 
\label{sec:SUSYscales}

In linear theory the collapse time of a typical spherical top-hat overdensity
of mass $M$ is given by $\sigma(M)D(a)=\delta_c$, where $D(a)$ is 
the growth factor and $\delta_c$ is the critical 
overdensity for collapse. Assuming a flat universe with $\Omega_M=1$ at these
early epochs, a typical $10^{-6}\msun$ microhalo will form at $z=30$, whereas 
a much more massive $0.01\msun$ parent will collapse at $z=23$, 
i.e. structure formation on very different mass scales happens almost simultaneously. 
By contrast, $\sigma(M)$ is significantly steeper on galaxy and 
cluster scales, which leads to a orderly hierarchical buildup of cosmic structure.

\begin{figure*}[ht]
\begin{center}
\includegraphics[height=8in]{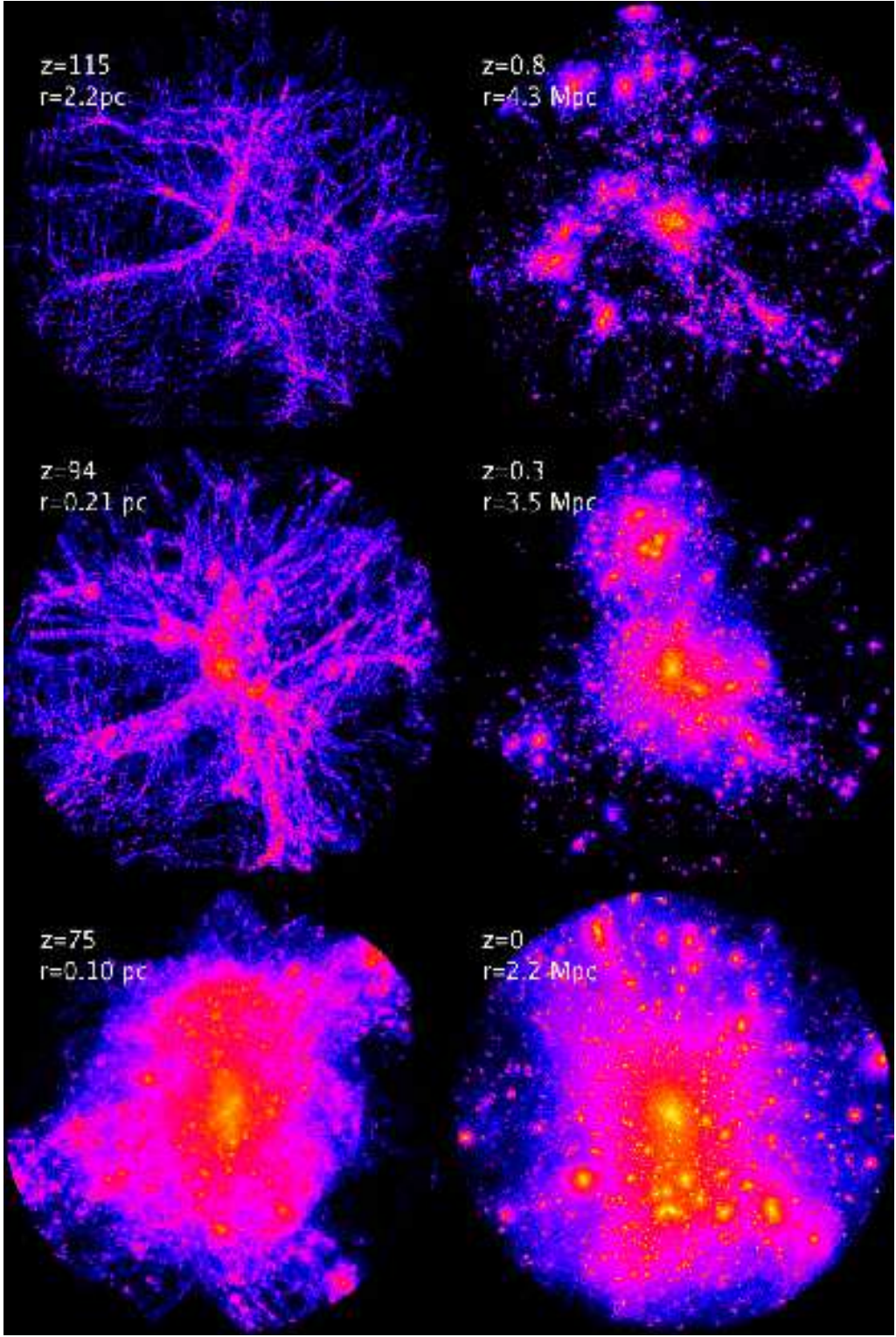}
\vspace{0.4cm}
\caption{Local DM density maps. The left-hand panels illustrate the almost
simultaneous structure formation in the SUSY run at different epochs,
within a sphere of physical radius $r$ including a mass of $0.014\,\msun$ 
(equal to $\mvir$ at $z=75$). The galaxy cluster halo B9 ({\it right-hand panels}) 
forms in the standard hierarchical fashion: the DM distribution is shown within 
a sphere of radius $r$ including a mass of $5.9\times 10^{14}\,\msun$
(equal to $\mvir$ at $z=0$). The SUSY and cluster halos have concentration 
parameter [defined as $c=2.15(\rvir/\rvcmax)$, which is equivalent to $c=\rvir/r_s$ for
a Navarro-Frenk-White (1997, hereafter NFW) profile] $c=3.7$ and $c=3.5$, 
respectively. In each image the logarithmic color scale ranges from 10 to 
$10^6\rho_{\rm crit}(z)$.
}
\label{fig:sixpanel}
\end{center}\end{figure*}

\begin{figure*}[ht]
\begin{center}
\includegraphics[width=3in]{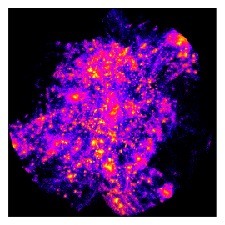}
\includegraphics[width=3in]{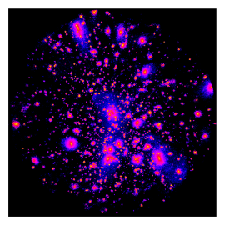}
\caption{Phase-space density map for the $z=75$ SUSY halo ({\it left}) and
the $z=0$ galaxy cluster ({\it right}). Local phase-space densities are estimated
using the SMOOTH program (available at www-hpcc.astro.washington.edu/tools/smooth.html) 
over 32 nearest neighbors. Note the different color scales: relative to the average
virial phase-space density ($\bar{\rho}/\bar{\sigma}_{\rm 1D}^3$), the logarithmic color
scale ranges from 10 to $10^{5}$ in the SUSY halo and from 10 to $10^7$ in the cluster
halo.}
\label{fig:phasespace}
\end{center}\end{figure*}

The sequence of nested N-body resimulations of \citet{Gao2005}
already showed that large-scale structure at $z\sim 50$ is qualitatively
different from that in the low-redshift universe. A comparison between
the extremely rapid assembly of our SUSY halo at redshifts $130<z<75$ and
the canonical hierarchical growth of cluster B9 at late epochs is 
given in Figure \ref{fig:sixpanel}. Here we show maps of the DM distribution 
at different epochs within a sphere enclosing the final virial mass of 
the parent halo, using identical (logarithmic) color scales for
direct comparison. The relative sizes of the most massive progenitors
are obviously much smaller for the SUSY host than for the cluster.
The virial mass of the SUSY halo grows by a much larger factor than B9 
during the comparable fractional increase in scale factor shown 
here. Another consequence of nearly
simultaneous structure formation is that filaments are very large
relative to the size of the biggest virialised systems in the same
region \citep{Gao2005}.

The difference in formation histories discussed above should be
reflected in the density contrast between substructure and the host,
since the average density at virialization is proportional to the mean
density of the universe at the time of collapse. The density contrast 
scales approximately as the inverse cube of the scale factor at collapse, 
$(a_{\rm halo}/a_{\rm subh})^3$. For early microhalos of masses 
$10^{-6} \msun$ and $10^{-2} \msun$ this ratio is only 2.5, much smaller 
than the typical contrast $\sim 20$ of galaxy-scale substructure in a cluster 
host. The expectation is then that substructure in B9 will have significantly 
higher contrast than in the SUSY halo\footnote{This simple argument is strictly
valid only for typical $1\sigma$ fluctuations of the density field. Subhalos are 
likely to have a slightly larger contrast, since they form from rarer fluctuations 
that are superposed on their parent halo density peak.}.
Visual inspection of Figure \ref{fig:sixpanel} clearly reveals this effect.

Both the SUSY run at $z=115$ and the B9 run at $z=0.8$ correspond to epochs
when the linear growth factor is about 0.66 of the final value at $z=75$ and
$z=0$, respectively. At $z=115$ we find 1931 virialized microhalos, 
less than the 2906 galaxy-size halos and subhalos in the cluster forming 
region at $z=0.8$. The abundance of bound halos in the SUSY run 
becomes similar to B9 only around $z\approx 100$. The fractional size of 
the most massive progenitor is also smaller in the SUSY run: at $z=115$ 
this has a mass of $0.01\,\mvir(z=75)$, while the largest
cluster progenitor at $z=0.8$ has already a mass of $0.25\,\mvir(z=0)$.
It takes until $z\approx 90$ before the SUSY main progenitor has also grown
to a virial mass of about one quarter of the final system. 
This illustrates the compression of host- and subhalo formation times
resulting from the flatness of $\sigma(M)$ on extremely small scales. 

Subhalos have smaller internal velocity dispersions than the hot
uniform density background. Therefore subhalos stand out much more
clearly in phase-space density $\rho/\sigma_{\rm 1D}^3$ than in real
space density, as shown in Figure \ref{fig:phasespace}. The phase-space map
reveals that the SUSY halo indeed has rich substructure at $z=75$,
comparable in abundance to the $z=0$ B9 cluster. Yet as in physical
density, the phase-space density contrast is lower in the SUSY run compared 
to the B9 halo. Note the different color scales
in Figure \ref{fig:phasespace}: the scale spans six decades in the B9
image, but only four decades in the SUSY image.

\subsection{Substructure mass function}
\label{sec:massfunction}

We have constructed substructure catalogs for the SUSY and B9 hosts at various 
stages in their formation. We employ two different algorithms to identify subhalos:
a) SKID \citep{Stadel2001}, which gives reliable subhalo masses
by iteratively removing unbound particles, but is computationally expensive; and b) a new
phase-space friends-of-friends method (6DFOF), which only produces
halo centers and maximum circular velocities, but is significantly
faster than SKID and is designed to take advantage of the higher
substructure contrast in phase-space. Both techniques are described in
more detail in Appendix \ref{sec:appendixA}. Because of the reduced
contrast at high redshift, the SUSY sub-microhalo catalogs depend strongly
on our choice of identification method and parameters, much more so
than in a $z=0$ cluster like B9. Nevertheless we find reasonable
agreement between the two methods when SKID is run with a linking
length of $0.0216$ comoving pc and a smoothing scale of $1024$ nearest
neighbors (cf. Appendix \ref{sec:appendixA}). 

\begin{figure}[ht]
\includegraphics[width=\columnwidth]{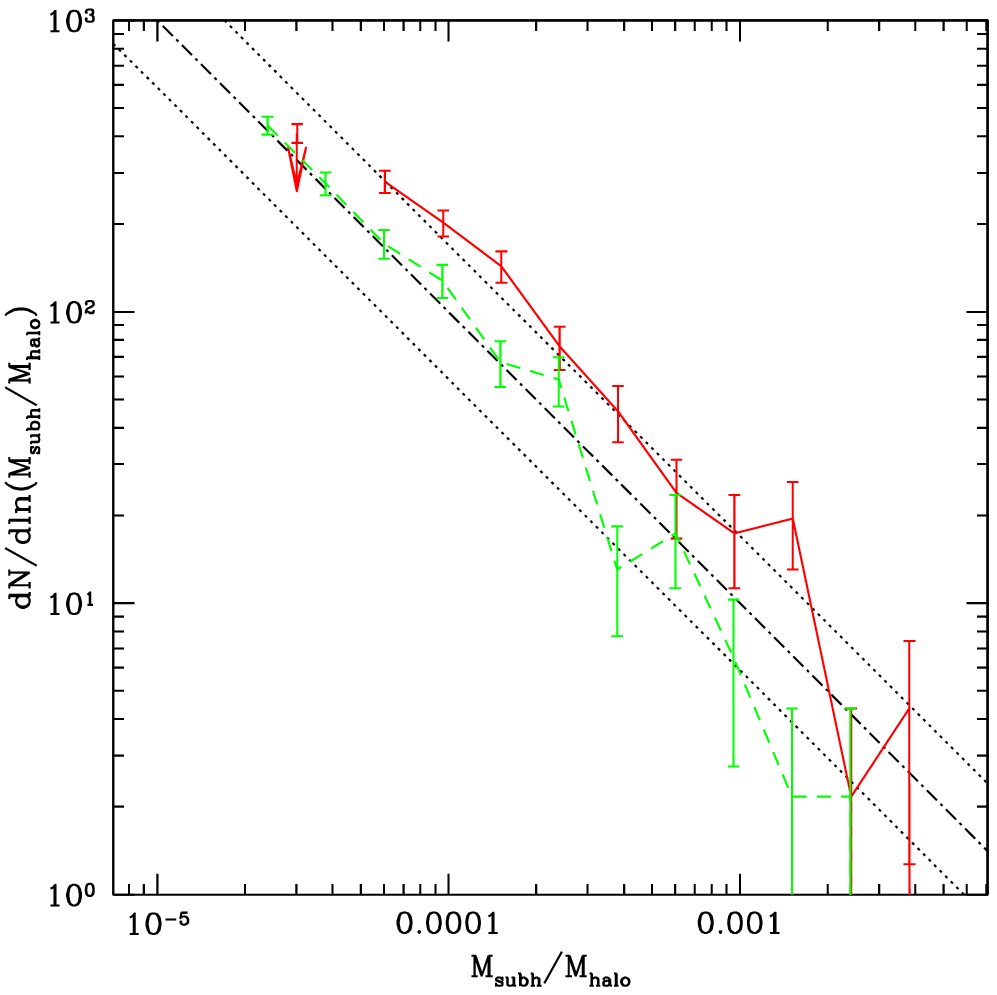}
\caption{Differential SUSY sub-microhalo abundance as a function of scaled 
sub-microhalo mass, $\msub/\mvir$, at $z=86$ ({\it solid}) and
at $z=75$ ({\it dashed}). Halo masses were determined by SKID using $s=1024$. 
The dash-dotted line corresponds to $dN/d\ln(\msub/\mvir)=0.01
(\msub/\mvir)^{-1}$, which is a good approximation to the $z=0$ galaxy
cluster relation \citep{Diemand2004sub}. The dotted lines show the 
scatter ($\approx 0.23$ dex) around this relation. The first bin
begins at 256 bound particles. Bars show the Poisson errors. The 
arrow shows the upper limit for the abundance of sub-$M_c$ clumps
(128 to 202 bound particles)
at $z=86$ estimated by using SKID with $s=64$ (see Appendix for details).
}
\label{fig:massfunction}
\end{figure}

\begin{figure}[ht]
\includegraphics[width=\columnwidth]{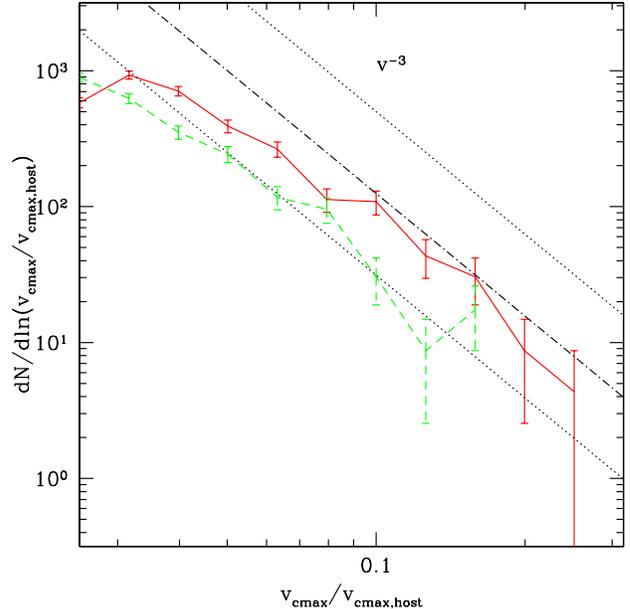}
\caption{Differential SUSY sub-microhalo abundance as a function of scaled 
maximum circular velocity, $\vcmax/\vcmaxhost$, at $z=86$ ({\it solid})
and $z=75$ ({\it dashed}). Halos were identified with a phase-space friend-of-friend 
finder (6DFOF). The dash-dotted line is the universal normalized  
circular velocity function of substructure in hosts of $10{11}$ to 
$10^{14}\,\msun$ found by \citet{Reed2005} at $0<z<2$, 
$dN/d(\vcmax/\vcmaxhost)=(1/8)(\vcmax/\vcmaxhost)^{-4}$.
The reported halo-to-halo variance in
the scatter around the universal circular velocity distribution is a factor
of 2 to 4, the dotted lines illustrate deviations by factors of four.} 
\label{fig:velocityfunction}
\end{figure}

We define a dimensionless subhalo mass, $\msub/\mvir$,
and plot in Figure \ref{fig:massfunction} the abundance of SUSY sub-microhalos 
at $z=86$ and $z=75$ against this normalized mass. The normalized mass function 
lies close to and within the scatter of the average normalized subhalo 
mass function found in galaxy clusters at the present epoch, a self-similarity that has
been previously found to hold over a wide range of mass scales and redshifts, 
from present-day clusters to dwarf galaxy halos at $z=4$ \citep{Moore2001}, and even
for $10^5$ $\msun$ host minihalos at $z=49$ \citep{Gao2005}.
Our simulations extend this range down to $10^{-2}\,\msun$ SUSY hosts at $z\gtrsim 
75$: the substructure populations of young, low-concentration halos are approximate scaled
copies of each other, independent of parent mass. Note that the same self-similarity 
also appears in the normalized distribution of subhalo maximum circular velocities 
(see Figure \ref{fig:velocityfunction}).

\begin{figure}[ht]
\includegraphics[width=\columnwidth]{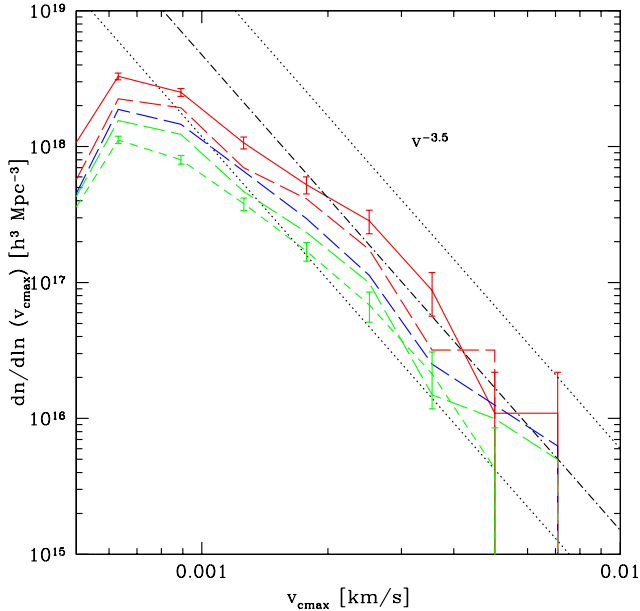}
\caption{SUSY sub-microhalo circular velocity function at $z=86$ ({\it solid
curve}), $z=83, z=80, z=78$ ({\it long-dashed curves}, from top to bottom),
and $z=75$ ({\it short-dashed curve}). Volume units are comoving.
The solid straight line is the universal subhalo velocity function 
of \citet{Reed2005}, extrapolated from $20-300\,\kms$ down to scales of ms$^{-1}$.
The dotted lines illustrate deviations by factors of four.}
\label{fig:pvelfabs}
\end{figure}

In Figure \ref{fig:pvelfabs} we present the non-normalized circular velocity
function for our set of SUSY sub-microhalos. The circular velocity distribution 
for subhalos of $20-300\,\kms$ in hosts of mass $10^{11}-10^{14}\,\msun$ at 
redshifts from zero to two is fitted by \citep{Reed2005}
\be
\label{eqn:pvelfabs}
\f{dn}{d\vcmax} = 1.5 \times 10^8 \vcmax^{-4.5} 
({\rm h}^3 {\rm Mpc}^{-3} {\rm km}^{-1} {\rm s})\;, 
\ee 
has little or no trend with host mass or redshift. The figure shows that our SUSY
sub-microhalo velocity function at $z=86$ lies right on the extrapolation (over four
orders of magnitude) of equation (\ref{eqn:pvelfabs}).
The sub-microhalo abundance appears to decrease with cosmic time until the last analyzed 
snapshot at $z=75$, where it is still within the scatter around the 
universal function of \citet{Reed2005}. Unfortunately we cannot
quantify the substructure abundance at lower redshifts since the parent
halo is growing very quickly and starts accreting material from the lower 
resolution regions, in the form of heavier particles. 
Note that the evolution in the subhalo mass or circular velocity functions 
should not be used to constrain the survival probability of substructure, since 
the mass of the host (and hence the volume probed) increases 
significantly with time.

\subsection{Formation of halos below the cutoff scale}
\label{sec:cutoff}

The SUSY-CDM scenario adopted here has an exponential cutoff in the
power spectrum $P(k)$ at a scale corresponding to a mass of
$M_c=10^{-6}\,\msun$. Below this mass the formation of structures is
suppressed relative to a CDM scenario without such a cutoff
\citep{Diemand2005susy}. But the suppression is not total: some halos
and subhalos with masses well below $M_c$ do actually form.  The
formation, abundance and properties of halos below a cutoff scale has
been studied in detail in the context of warm dark matter (WDM)
models, both for field halos (e.g. \citealt{Bode2001};
\citealt{Knebe2003}; \citealt{Yoshida2003}) and subhalos
(e.g. \citealt{Colin2000}; \citealt{Yoshida2003}). A quantitative
comparison with these results lies beyond the scope of this work. In
this section we illustrate one formation path of sub-$M_c$ halos
and place an upper limit on their abundances in order to ensure that
DM clumps below $M_c$ do not provide a significant contribution to the
DM annihilation signal.

\begin{figure*}[ht]
\begin{center}
\includegraphics[height=8in]{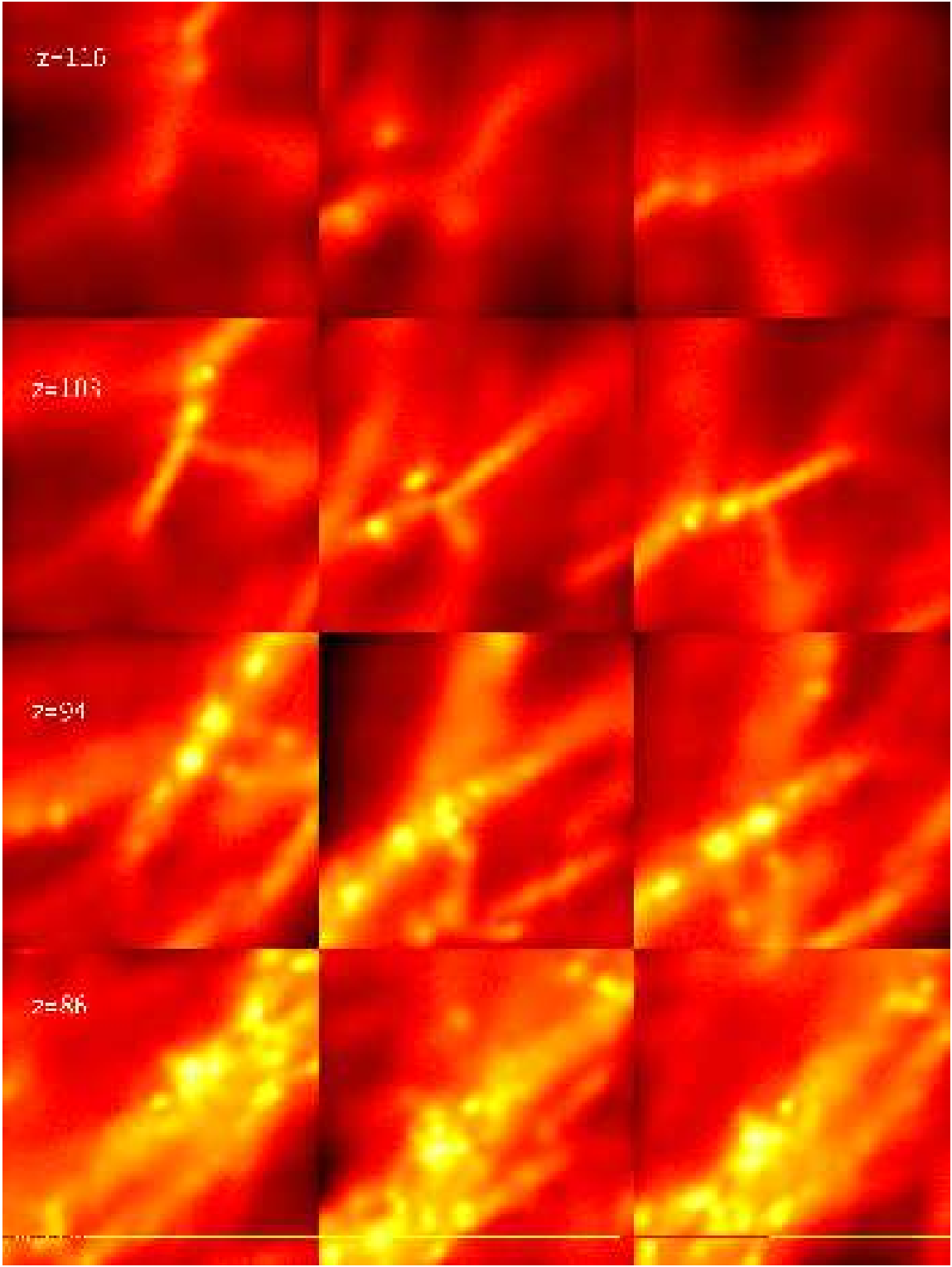}
\vspace{0.4cm}
\caption{Projected DM density maps illustrating the formation
of a halo with $\mvir=0.3\,\times 10^{-6}\,\msun$, i.e. below the
cutoff scale of $M_c = 10^{-6}\,\msun$. Three orthogonal projections of the 
same cube are shown at each time. The cubes have a size of
$L = 2.7$ comoving pc and are centered on those particles which end up in 
the sub-$M_c$ clump at $z=86$. The logarithmic color-scale goes from 
$\rho_{\rm crit}(z)\;L$ to $10^{2.4} \rho_{\rm crit}(z)\;L$ at all times. 
The whole region is relatively overdense: its mean density increases from 
$6 \rho_{\rm crit}(z)$ at $z=116$ to $21 \rho_{\rm crit}(z)$ at $z=86$.
The three projection directions are aligned with the axes of the initial 
grid and the initial particle spacing of 0.3 comoving pc is indicated by 
the ruler in the lower-right panel. Note that these structures do not show 
a preferred direction, alignement with the axes of the initial grid, or 
any regular, grid-scale separation which would indicate
artificial fragmentation caused by the grid initial conditions.
}
\label{fig:cutoff}
\end{center}\end{figure*}

The perturbed cubic grid initial conditions used here have a 
characteristic length scale and preferred directions, which is not the case for 
unordered ``glass'' initial conditions \citep{White1993}. \cite{Goetz} suggested
that grid initial conditions lead to artificial 
fragmentation of WDM filaments and that glass initial conditions should be used 
instead to overcome this problem. Even glass WDM simulations show the formation
of sub-cutoff objects \citep{Yoshida2003}, however, meaning either that glass 
initial conditions do not prevent artificial fragmentation or that
the formation of sub-cutoff structure is real. Our sub-$M_c$ halos do not form
with a preferred separation or along the initial grid axis 
(see Figs. \ref{fig:sixpanel} and \ref{fig:cutoff}), which implies that 
they are not artifacts produced by regular grid noise in the initial conditions.

Figure \ref{fig:cutoff} shows the formation of a clump of mass $0.3\times
10^{-6}\,\msun$ at $z=86$. This halo forms `top-down' through the fragmentation
of a filament, like the smallest WDM halos seen in \citet{Knebe2003}. Note 
that the fragmentation occurs only after the formation of larger halos in 
the same region, i.e. the sub-$M_c$ clump forms after its larger neighbors and
it therefore has a lower peak density and concentration. One can think of the 
sub-$M_c$ clump as a low amplitude initial overdensity on some scale larger than $M_c$
which fails to grow to a mass above $M_c$: by the time the low amplitude 
perturbation goes non-linear its larger and higher amplitude neighbors
have already collapsed and are accreting some of the mass which in linear theory 
would correspond to the smaller, low amplitude peak. The small peak collapses at
a later time in a relatively depleted environment and fails to grow above $M_c$.

Structures above $M_c=10^{-6}\,\msun$ from almost at the same time
(see section \ref{sec:SUSYscales}) and those below $M_c$ appear to form even later.
This explains the low concentration and density contrast of sub-$M_c$ clumps. 
Their measured abundance depends strongly on the operational definition of 
substructure, i.e. on the subhalo finding algorithm used and its parameter 
values (see Appendix for details).
The largest abundance of sub-$M_c$ clumps is found with SKID using a
density estimate over 64 particles (see Figure \ref{fig:massfdep}). This abundance 
must be taken with caution, however, since many of the smaller groups identified 
in this way do not show up as significant peaks in local density or phase-space 
density maps. We use the $s=64$ SKID value only as an upper limit to the number
sub-$M_c$ clumps. It demonstrates that the subhalo mass function becomes 
much shallower than $dn/d\msub\propto \msub^{-2}$ below about $0.5\,M_c$ 
(Figure \ref{fig:massfunction}).
Together with their late formation times and low concentration this is
sufficient to exclude a significant
contribution to the DM annihilation signal (see section \ref{sec:annihilation})
from clumps below $M_c$.

\subsection{Survival}
\label{sec:survival}

In order to properly quantify the survival probabilities and
robustness against tidal stripping of subhalos at high and low
redshift we have traced substructure in the entire high resolution region
forward in time. We examine three time intervals, each covering a
comparable increase in scale factor. Because of the rapid assembly of the 
SUSY halo we are limited to a relatively small ratio of final to initial
scale factor, $a_f/a_i \approx 4/3$, i.e. from $z_i=115$ to $z_f=86$ and 
from $z_i=101$ to $z_f=75$ for the SUSY halo, and from $z_i=0.333$ to the
present epoch for the B9 cluster. Subhalos are identified with 6DFOF 
(see Appendix \ref{sec:appendixA}) and we consider only halos with more 
than 320 linked particles at $z_i$, an order of magnitude
larger than our usual choice of 32 linked particles at $z_f$. This allows 
us to identify surviving structures that have lost a large fraction of their 
mass (up to 90 percent) due to tidal effects. 
A halo at $z_i$ is counted as a survivor if it contributes at least 10
particles to some halo at $z_f$, unless it has merged with the central
core of the $z_f$ parent system. Note that even halos for which we find no remnant 
may actually have one at higher numerical resolution.

\begin{figure}[ht]
\includegraphics[width=\columnwidth]{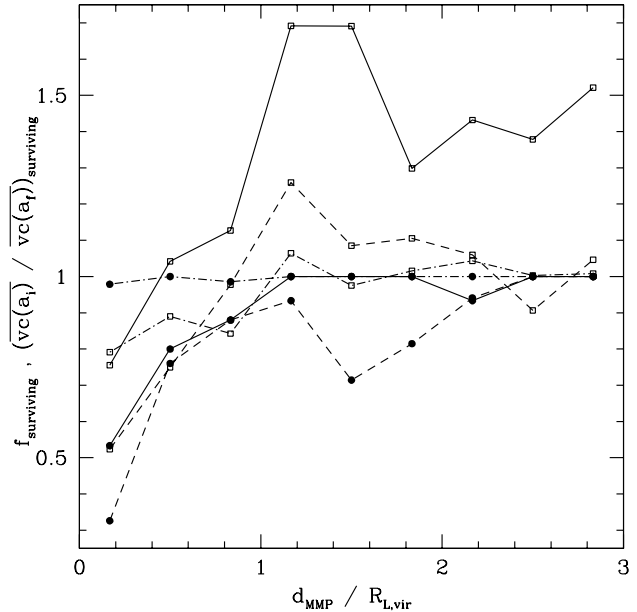}
\caption{Fraction of surviving structures ({\it solid circles}) and ratios
of initial to final average circular velocity $\vcmax$ of surviving 
substructure ({\it open squares}) versus initial separation from the most 
massive progenitor in units of $\rlagrange$. 
Halos were traced from $z_i=115$ down to $z_f=86$ ({\it solid lines}), 
from $z_i=101$ to $z_f=75$ ({\it dashed}), and from $z_i=0.333$ down to 
the present epoch ({\it dash-dotted}).}
\label{fig:vcs}
\end{figure}

We approximate the Lagrangian region of the particles that will end
up within the virial radius of the parent at $z_f$ by a sphere of radius
$\rlagrange$ centered on the most massive progenitor (MMP) at $z_i$, and 
enclosing a mass $\mvir(z_f)$. Halos within $\rlagrange$ at
$z_i$ are likely to end up within $\rvir$ at $z_f$. Table
\ref{tab:survial} summarizes the survival probabilities of well
resolved substructure as a large host system is forming. Considering all
halos in the high resolution region (top three rows) we see that early
substructure is significantly more likely to be disrupted, with more
than $99\%$ of all subhalos surviving from $z_i=0.333$ down to the present epoch,
but only $87\%$ surviving from $z_i=115$ to $z_f=86$ and even less ($73\%$) from
$z_i=101$ to $z_f=75$. Tidal destruction is more efficient closer to the
center of the host, and this causes a further reduction in the
survival fraction of substructure within two (middle three rows) or one
$\rlagrange$ (bottom three rows). In Figure \ref{fig:vcs} we plot the
survival fraction versus distance to the MMP in
units of $\rlagrange$, $d_{\rm MMP}$, for the three time epochs analyzed. This
drops rapidly for $d_{\rm MMP}\lesssim \rlagrange$ in the case of SUSY sub-microhalos, below
$50\%$ ($30\%$) for the epoch ending at $z_f=86$ ($z_f=75$).

The subhalos that do survive experience significant mass loss due to
tidal stripping. We quantify this mass loss by calculating the ratio
of the average of $\vcmax$ of all halos at the beginning to that at
the end of each epoch, $F_{\vcmax} \equiv \langle v_{\rm
c,max,i}\rangle/\langle v_{\rm c,max,f}\rangle$ (Figure \ref{fig:vcs}). For 
halos beyond $\rlagrange$ $\vcmax$ grows with time in the SUSY simulation, whereas for
B9 the ratio remains about constant. The increase
in average $\vcmax$ is especially strong in the redshift interval $115-86$, 
$F_{\vcmax} \approx 1.5$, and this again illustrates the rapid and
late (in terms of $a_f/a_i$) formation and growth of these
structures. SUSY sub-microhalos within $\rlagrange$ suffer
more mass loss than those beyond $\rlagrange$,
$F_{\vcmax}(<\rlagrange)=0.68F_{\vcmax}(>\rlagrange)$. For the cluster
substructure the difference between outer and inner regions is less
pronounced, indicating that it is more resistant against tidal
stripping. This robustness translates into a higher survival
probability, as shown above.

\begin{table}
\centering
\caption{Numbers and fractions of surviving structures}
\label{tab:survial}
\begin{tabular}{l|ccccccc}
\hline
Region& $z_i$ &$z_f$&$N_{\rm tot}$&$N_{\rm sur}$&$f_{\rm sur}$&$N_{\rm core}$& $F_{\vcmax}$\\
\hline
all & 115 & 86 & 150 & 131 & 0.873 & 3 & 1.286 \\
    & 101 & 75 & 227 & 165 & 0.727 & 2 & 0.954 \\
    & 0.333& 0 & 774 & 771 & 0.996 & 1 & 0.956 \\
\hline
$<2\rlagrange$
& 115 & 86 & 111 & 94 & 0.847 & 3 & 1.242 \\
& 101 & 75 & 206 & 145& 0.704 & 2 & 0.944 \\
& 0.333& 0 & 497 & 494& 0.994 & 1 & 0.925 \\
\hline
$<\rlagrange$
& 115 & 86 & 75 & 58 & 0.773 & 3 & 1.033 \\
& 101 & 75 & 121& 75 & 0.620 & 2 & 0.765 \\
& 0.333& 0 & 261& 258& 0.989 & 1 & 0.8414 
\end{tabular}
\tablecomments{Substructure is identified based on phase-space density
(using 6DFOF) at $z_i$ and its survival is checked at $z_f$. $N_{\rm
tot}$ and $N_{\rm sur}$ are the total number and number of surviving halos, 
and $f_{\rm sur}=N_{\rm sur}/N_{\rm tot}$ denotes the fraction of
surviving substructure. $N_{\rm core}$ is the number of halos merging
with the core of the $z_f$ halo. $F_{\vcmax} =\langle v_{\rm
c,max,i}\rangle/\langle v_{\rm c,max,f}\rangle$ denotes the ratio of the
initial to final average maximum circular velocity. The first group of
three rows is for the entire high resolution regions, the second and
third groups are for substructures initially within two and one
$\rlagrange$, respectively, where $\rlagrange$ is the radius
encompassing $\mvir(z_f)$ at $z_i$. See text for details on the 6DFOF
phase-space halo finder and on how survival is determined.
\\
}
\end{table}

\section{Dark matter annihilation signal}
\label{sec:annihilation}

The differences in substructure contrast described in the previous
sections have ramifications for the expected $\gamma$-ray signal from
DM self-annihilation. Indirect detection experiments are
hoping to detect both the cosmological extragalactic background as
well as the local signal from the Milky Way's center and from DM 
substructure in its halo.  Since the annihilation rate is proportional
to $\rho_{\rm DM}^2$, the predicted flux depends sensitively on the
clumpiness of the DM distribution. Small-scale substructure is 
expected to significantly increase the signal from individual halos
\citep{CalcaneoRoldan2000,Stoehr2003,Colafrancesco2005}, as well as the
the diffuse extragalactic background
\citep{Ullio2002,Taylor2002,Ando2005}. Annihilation luminosity estimates
based on DM densities measured in collisionless N-body
simulations of a Milky Way-scale halo suggest that 
substructure may boost the signal by only a small factor, $\sim 70\%$,
compared to a smooth spherical NFW halo (Stoehr et al. 2003, 
hereafter S03). In this work we follow S03's analysis and compare
the contribution of substructure in the SUSY host at $z=75$ and
the B9 galaxy cluster at $z=0$.

The annihilation luminosity is proportional to the ratio of the
velocity weighted self-annihilation cross section and the mass of the
DM particle, $\langle \sigma v\rangle/m_X$. Following S03, we
disregard this (highly uncertain) particle physics term and concentrate
instead on the astrophysical component, defining the annihilation
signal of the $i^{\rm th}$ halo to be
\be
S_i = \int_{V_i} \rho^2_{\rm DM} dV_i = \sum_{j \epsilon \{P_i\}} \rho_j m_j, 
\label{eq:annihilation}
\ee
where $\rho_j$ and $m_j$ are the density and mass of the $j^{\rm th}$
particle, and $\{P_i\}$ is the set of all particles belonging to halo
$i$. Whereas S03 estimated the $\rho_j$ with the
parameter-free Voronoi tessellation technique, we use the SPH-kernel
estimated densities from SMOOTH over 64 nearest neighbors.

Like S03 we find that the component from the spherical-average density
profile contributes only a fraction of the total signal, $49\%$ in the
SUSY halo and only $35\%$ in the B9 halo. The contribution to the
signal from all resolved bound subhalos is $39\%$ in B9 
and only $10\%$ in the SUSY host. The remaining difference is due to both
unbound streams and departures from spherical
symmetry\footnote{These estimates depend only weakly on the smoothing scale
used for the SPH density estimates: using 1024 instead of 64 
nearest neighbors gives  
relative contributions of spherical smooth signal, bound substructure, and
remaining density fluctuations of $(57\%, 9\%, 34\%)$
instead of $(49\%, 10\%, 40\%)$ for the $z=75$ SUSY halo.}.
The absolute contribution of the resolved 
substructure component depends on numerical resolution, since higher
resolution simulations would be able to resolve higher densities in
the subhalo centers \citep{Kazantzidis2004}. 
The bound substructure contribution to the total signal could
be larger and the values given here should be taken as lower limits. Both 
the SUSY and B9 halos, however, have approximately the same number of 
particles within their virial radii, allowing us to directly compare 
the contribution from resolved substructure between the two. We conclude 
that it is
indeed the lower density contrast in the early SUSY host that leads to a
significantly smaller contribution from bound sub-microhalos to the total 
annihilation signal.

\begin{figure}[ht]
\includegraphics[width=\columnwidth]{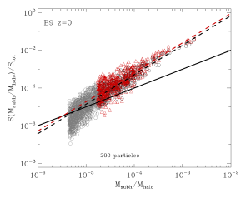}
\includegraphics[width=\columnwidth]{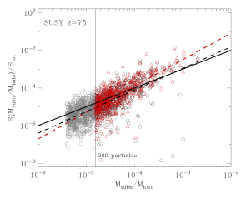}
\caption{
The ratio of the annihilation signal of individual subhalos to the
total signal from all dark matter particles in the host halo versus
the subhalo mass ratio $\msub/\mvir$. Circles denote the signal
calculated directly from the subhalo particle densities, triangles are
used for an estimate of the signal assuming an NFW profile and a
concentration $c \approx 2 \rvir/\rvcmax$. The top panel shows the B9
halo at $z=0$, the bottom one the SUSY halo at $z=75$. Subhalos with
the same signal-to-mass ratio as the host would lie on the thick solid
line. The dashed lines show the best-fit linear relation between
subhalo signal and normalized mass (\textit{lower line:} direct
calculation, \textit{upper line:} NFW estimate). Halos to the right of
the thin vertical line have more than 200 particles.
}
\vspace*{0.1in}
\label{fig:subsignal}
\end{figure}

In Figure \ref{fig:subsignal} we present the mass dependence of the
annihilation signal for individual subhalos. To ease a direct
comparison we normalize the subhalo mass by the total halo mass and
the subhalo signal by the total signal from all halo DM 
particles. Since the signal strength may depend sensitively on the
internal density structure of the subhalo, we only include halos with
more than 200 particles in our analysis. If substructure had the same
signal-to-mass ratio $S/M$ as their host they would lie on the
thick solid line in the figure. We find instead that the B9 subhalos 
tend to have a larger $S/M$ (i.e. they lie above the solid line, in 
qualitative agreement with S03), but that the majority of the
SUSY sub-microhalos have lower $S/M$ than their host. Furthermore we
find that $S/M$ is not independent of mass: in both our simulations
less massive substructures have a lower $S/M$ than more massive ones. For
B9 we find $S/M \propto \msub^{0.5}$, while for SUSY $S/M\propto \msub^{0.4}$. 
Given a substructure abundance of $dN/d\msub\propto \msub^{-\alpha}$ with $\alpha 
\lesssim 2$ (see Section \ref{sec:massfunction}), this suggests that the total signal may be
dominated by the highest mass subhalos. One has to keep in mind, however,
that the peak densities in an N-body halo are always limited by
numerical resolution. The local densities are too low both in the
center of the host system \citep{Diemand2004cluster} and even more so
in the inner regions of subhalos \citep{Kazantzidis2004}. It is thus
likely that estimates of the total annihilation signal based directly
on densities measured in N-body halos underestimate the importance of
the substructure contribution. Analytic calculations 
\citep{CalcaneoRoldan2000} tend to find larger substructure boost
factors and a signal that is dominated by small subhalos
\citep{Colafrancesco2005}.

We have attempted to correct for the artificial heating of the densest,
inner regions by recalculating the annihilation signal for every
subhalo assuming an NFW profile with a concentration given by $c
\approx 2.16 \rvir/\rvcmax$. In this case the normalized annihilation
signal is equal to
\be \label{eq:NFWsignal}
\tilde{S}/S_{\rm tot} \approx 1.22 \left(\f{\vcmax^4}{G^2 \rvcmax}\right)/S_{\rm tot},
\ee
where $G$ is the gravitational constant, and $\vcmax$ and $\rvcmax$
are determined using SKID with spherical binning of the bound subhalo
particles. Expressing $\tilde{S}_i$ in terms of $\vcmax$ and $\rvcmax$
reduces the dependence of the estimate on the strongly resolution
dependent inner density profile. These corrected subhalo signals are
plotted as triangles in Figure \ref{fig:subsignal}. 
As expected the NFW estimate enhances the signals, and the correction
is about $30\%$ in both the SUSY and the B9 halo. With this boost the relative
contributions of spherical smooth signal, bound substructure, and
remaining density fluctuations are $(47\%, 14\%, 39\%)$ for the
SUSY and $(30\%, 48\%, 22\%)$ for the B9 halo. Note that the substructure
signal in the B9 run is still significantly larger than in the SUSY
run. The mass dependence of $S/M$ appears to be preserved, but we caution that
even our NFW estimates are not bias free: the finite resolution in our
simulations may well lead to an underestimate of $\vcmax$
\citep[cf.][]{Kazantzidis2004}, especially for low-mass subhalos.

Our results suggest that in a real sub-solar mass SUSY halos the 
signal from subhalos and other density fluctuations is at least as 
big as the signal from the spherically-averaged
component.  Estimates of the cosmological, extragalactic $\gamma$-ray
background \citep{Ullio2002,Taylor2002,Ando2005} should be revised to
take into account both microhalos in the field \citep{Diemand2005susy}
and sub-microhalos within larger (sub)-solar mass systems like the one
studied in this work. We plan to address remaining open questions
(i.e. the total substructure boost factor and what subhalo mass scale
dominates the signal) in future work using higher resolution
simulations, extensive convergence tests, and comparisons with various
analytical estimates.

\section{Conclusions}
\label{sec:conclusions}

We have simulated the formation of a $0.014\,\msun$ halo at $z=75$ in
a scenario where CDM consists of 100 GeV supersymmetric 
particles. Our main results are:
\begin{itemize}

\item Early SUSY-CDM structures form almost simultaneously over a wide
range of mass scales. Halo masses
grow very rapidly, and filaments are very extended compared
to the size of the largest halos.

\item The contrast between substructure and background is lower 
in early SUSY halos than in galaxy cluster hosts, both in density and
in phase-space density.

\item At $z=0$ a simple, fast phase-space friends-of-friends subhalo
finder (6DFOF) gives robust results. Velocity functions and subhalos
number density profiles are identical to those obtained with
SKID. Unfortunately on SUSY scales both methods have problems due
to the reduced contrast. Results from both methods depend
quite strongly on the input parameters, but can be brought into
agreement with a suitable choice of parameters.

\item In relative units both the sub-microhalo mass and circular velocity functions
of the $0.014\,\msun$ SUSY host are in the range found for $z=0$ galaxy clusters.

\item In absolute units both the circular velocity functions of the
$0.014\,\msun$ host halo lie within the scatter of the universal
subhalo circular velocity function found by \citet{Reed2005} in low redshift
dwarf, galaxy, and cluster halos.

\item As the universe expands by a factor of 1.3, about 20 to 40
percent of well-resolved early substructure is destroyed, compared to
only about 1 percent in the low-redshift cluster. The reductions in
the peak circular velocities among the survivors are also larger for
the early micro-subhalos.

\item The contribution to the DM annihilation signal from the
bound sub-microhalos resolved in our simulations is about four times smaller
than the signal from the spherically-averaged component in the SUSY
host. In a cluster simulated at similar numerical resolution (B9),
the bound substructure contribution exceeds the spherically-symmetric part.
Due to numerical limitations these are lower limits. The reduced contrast 
in early-forming sub-microhalos results in a lower substructure boost 
factor compared to galaxy or cluster halos.

\end{itemize}

\section*{Acknowledgments}

We would like to thank the referee for helpful comments.
It is a pleasure to thank Joachim Stadel for making PKDGRAV available 
to us and for his help with the implementation of the 6DFOF group-finder.
We thank Tom Abel, Andreas Faltenbacher, Ben Moore, Joel Primack and Simon 
White for useful and motivating discussions.
J.D. especially thanks Paul Schechter and Ed Bertschinger for 
discussions and suggestions during the MKI dark matter substructure workshop at
MIT. Support for this work was provided by NSF grant AST02-05738 and
NASA grant NNG04GK85G (P.M.), and by the Swiss National Science
Foundation (J.D.). Simulation B9 was run on the zBox supercomputer at
the University of Zurich, while the SUSY run was performed on NASA's
Project Columbia supercomputer system.

\pagebreak

\appendix

\section{Identification of early SUSY-CDM micro-subhalos}
\label{sec:appendixA}

The simultaneous nature of structure formation near the SUSY cutoff
scale at early times leads to substructure with a reduced contrast
both in real and phase-space density. Conventional subhalo finders use
positional information in the group finding stage and velocities only
after the fact to decide which particles are gravitationally bound to
a subhalo. In this appendix we discuss how a conventional subhalo
finder (SKID) performs in these challenging low contrast systems and
present a simple and fast new method of finding (sub-)structures in
phase-space.

\subsection{SKID: self bound over-densities}

SKID (publicly available at: www-hpcc.astro.washington.edu/tools/skid.html)
is described in detail in \citet{Stadel2001}. In a dark matter only
simulation the group-finding algorithm goes through the following
steps. First, the local densities are calculated using an compact SPH
kernel whose scale is adjusted for each particle to includes a fixed
number '$s$' of nearest neighbors (the same method is used in 
the program SMOOTH available at www-hpcc.astro.washington.edu/tools/smooth.html).
Second, particles are
moved along the density gradients until they move less than a user
specified length $\epsilon$, i.e. they are trapped in a local
potential well of scale $\epsilon$. When all particles satisfy this
criterion, they are linked together using a linking length equal to
$\epsilon$. Finally the gravitationally unbound particles are removed
iteratively (one by one) from all the groups.

SKID results depend on two parameters: the linking length $\epsilon$
and the number of nearest neighbors $s$. A small $\epsilon$ tends to
underestimate the mass of larger subhalos while a large $\epsilon$
links nearby subhalos together. In current high resolution $z=0$ CDM
clusters (roughly ten million particles within the virial radius)
setting $\epsilon =0.004\,\rvir$ successfully identifies most resolved
structures. Combining subhalo catalogs derived with a range of values
for $\epsilon$ results in small corrections at the low mass and
high mass end of the subhalo mass function
\citep{Ghigna2000,Diemand2004sub}. For the $z=86$ SUSY halo we find a
similar weak $\epsilon$-dependence of the SKID results.

\begin{figure}[ht]
\includegraphics[width=3.5in]{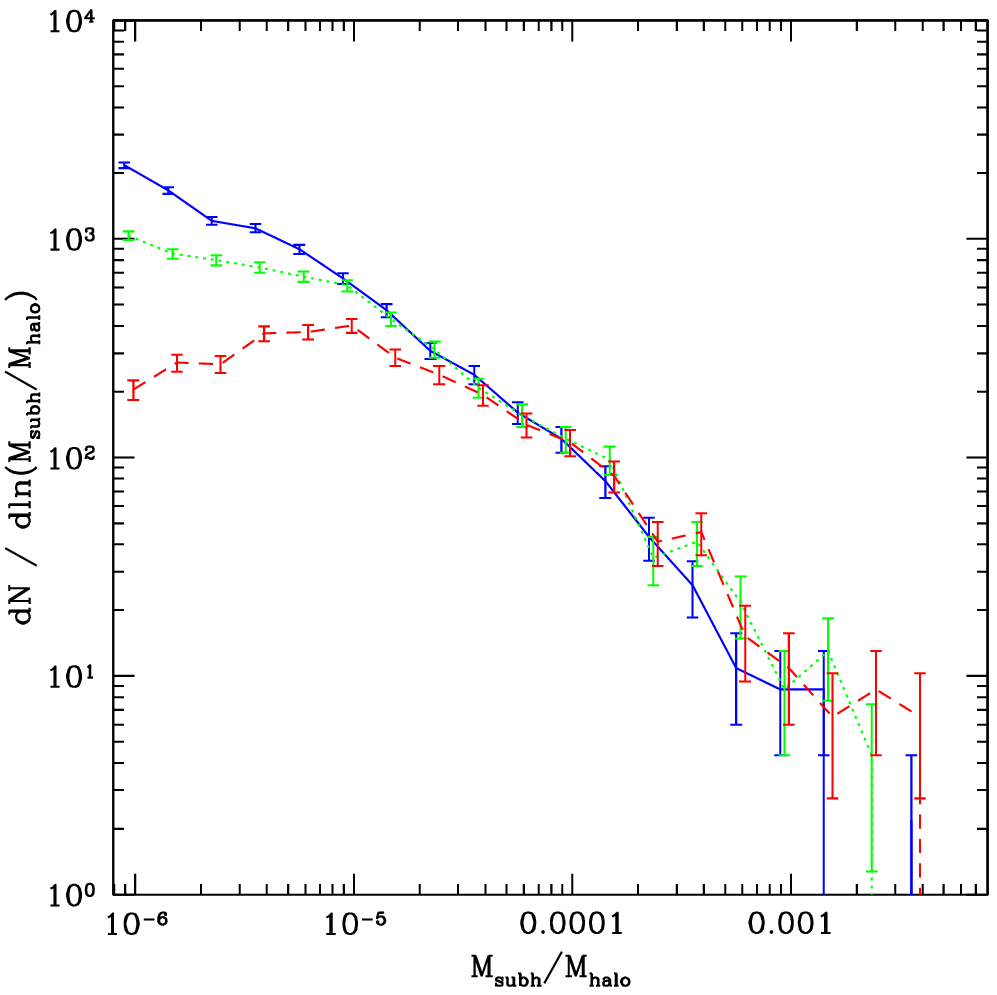}
\includegraphics[width=3.5in]{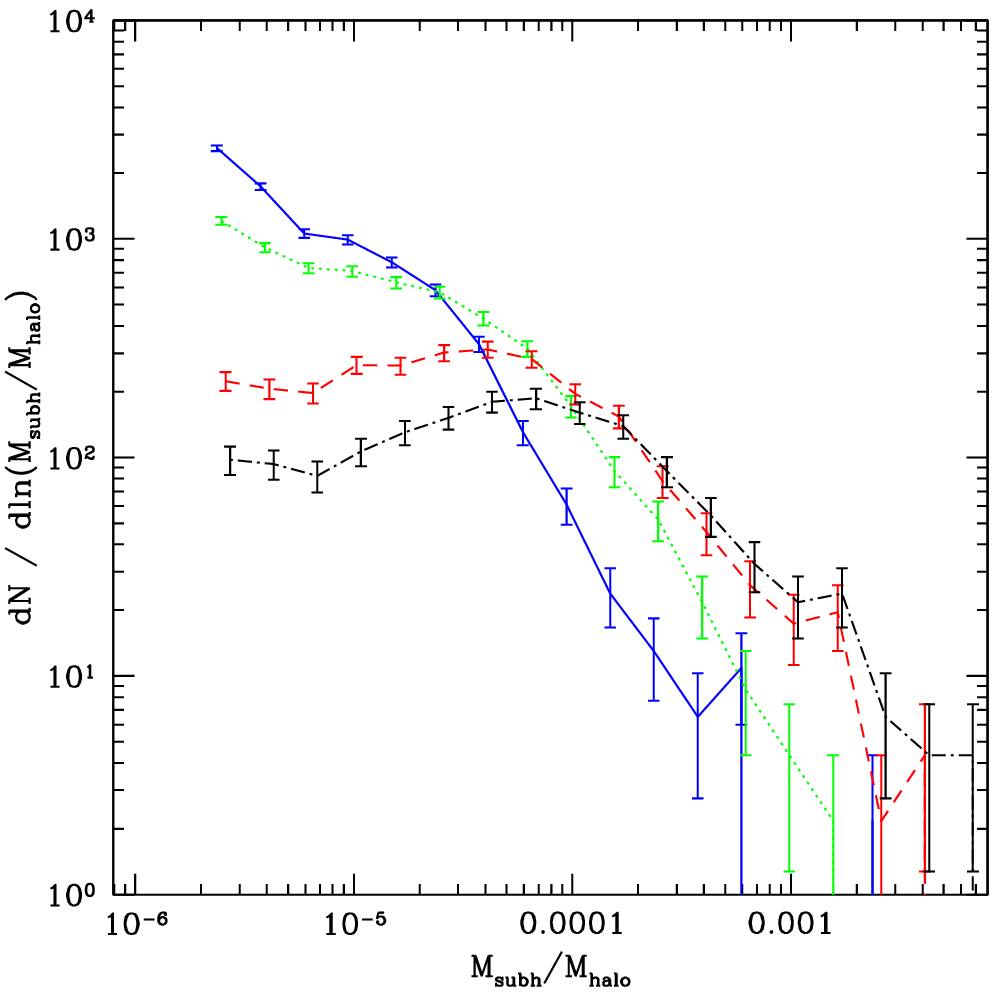}
\caption{SKID subhalo mass functions for the $z=0$ galaxy cluster
(\textit{left}) and the $z=86$ SUSY CDM halo (\textit{right}) using different numbers of
neighbors $s$ for the SPH density calculation. Plotted are $s=64$
(\textit{solid}), 256 (\textit{dotted}), 1024 (\textit{dashed}), and 2048 (\textit{dash-dotted},
SUSY halo only). The first bin always begins at 10 bound particles. Bars
show the Poisson errors.  For the massive halo, the mass functions are
independent of $s$ above about $0.25 s$. Due to the smaller density
contrast the entire mass function depends on $s$ for the microhalo.
}
\label{fig:massfdep}
\end{figure}

The number of nearest neighbors $s$ is usually set to $s=64$ and in a
$z=0$ CDM cluster the SKID subhalo mass and velocity functions are
practically independent of $s$, with the exception that some subhalos
with less than about $0.25 s$ bound particles are lost when a larger
smoothing scale is used (see Figures \ref{fig:massfdep} and
\ref{fig:pvelfdep}). Unfortunately SKID's results are much more
parameter dependent for the SUSY halo: the right panel of Figure
\ref{fig:massfdep} shows that the SUSY subhalo mass function depends
on $s$ over the entire subhalo mass range. Many of the small subhalos
found with $s=64$ and $s=256$ appear spurious, i.e. visual inspection
fails to identify a clear corresponding peak in the mass (or phase
space) density projections. Only at the high mass end there is hint of
convergence when using $s>1000$. Tests in the $z=0$ cluster
suggest that using large $s$ leads to an underestimates of the subhalo
abundance below about $0.25 s$ bound particles. We compromise and use
$s=1024$ throughout the paper, but caution the reader that the
resulting group catalogs are probably incomplete below about $256$
bound particles.

\subsection{Finding structures in phase-space: 6DFOF}
\label{sec:appendixA2}

Subhalos have smaller internal velocity dispersions $\sigma$ than the
hot uniform density background.  Figure \ref{fig:phasespace} show maps
of $\rho / \sigma^3$ (both $\rho$ and $\sigma$ are calculated using an
SPH kernel over 32 nearest neighbors), which is a simple and fast way
to estimate the local phase-space density. The phase-space density
contrast is much larger than the real-space density contrast. Even
small subhalos, which due to limited numerical resolution don't reach
peak mass densities as high as their host halo, stand out more clearly
in phase-space density because internal velocity dispersions are
smaller for less massive halos. In order to take advantage of the
increased contrast in phase-space we have developed a new, simple, and
fast (sub-)halo finder.

\begin{figure}[ht]
\includegraphics[width=\columnwidth]{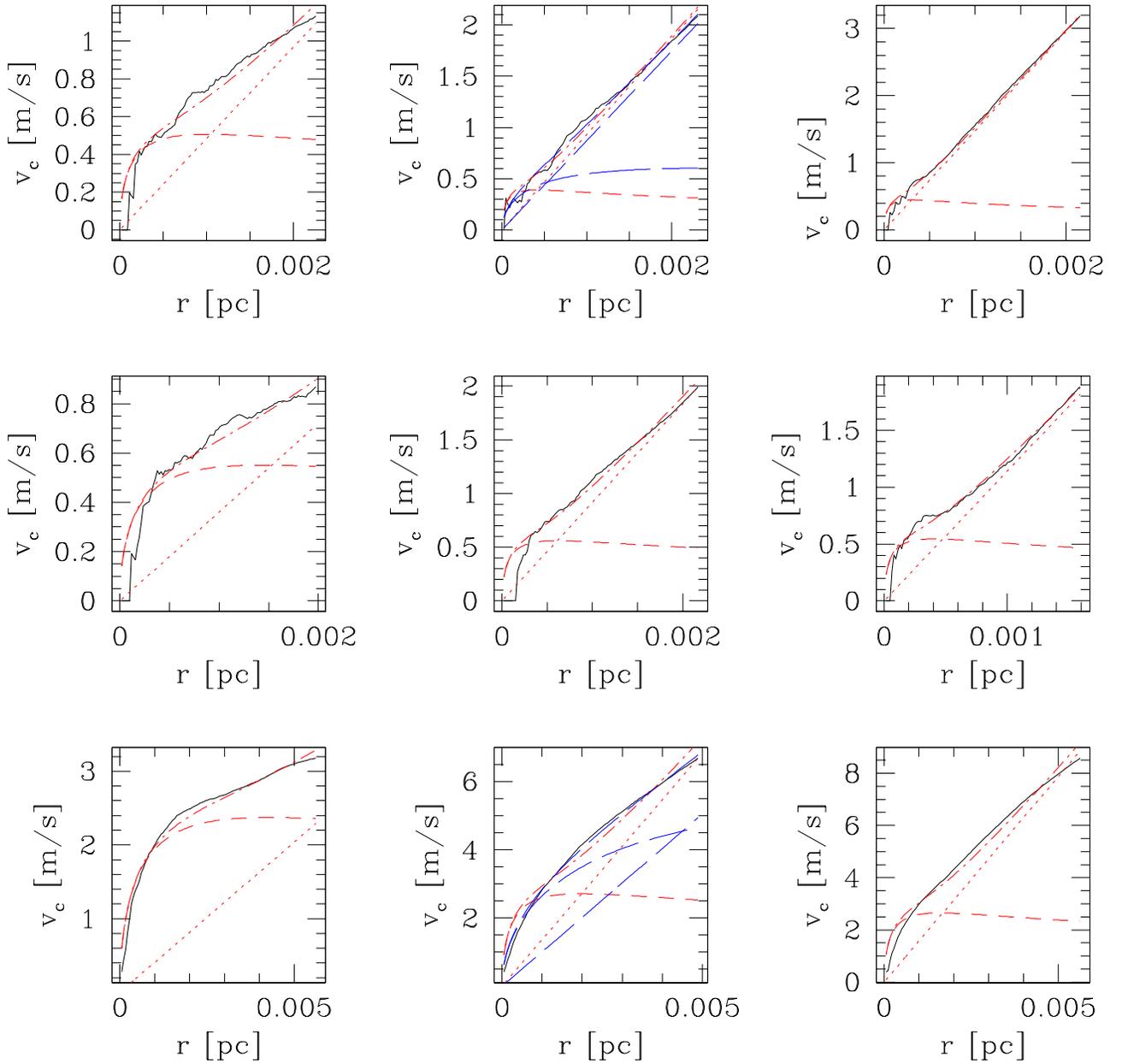}
\caption{Example circular velocity profiles (\textit{solid}) of structures
found with 6DFOF within the virial radius of the $z=86$ SUSY CDM halo. 
Best fit total circular velocity profiles $v_{c} (r)$
(\textit{dashed-dotted}) contain a contribution from an NFW subhalo
$v_{\rm c,sub}$ (\textit{dashed}) and from a homogeneous background
density $v_{\rm c,BG}$ (\textit{dotted}), i.e.  $v_{c} (r) =
\sqrt{v_{\rm c,sub}^2 + v_{\rm c,BG}^2 }$.  Upper panels show small
subhalos below the completeness limit of about $0.05 \vcmaxhost =
1\,\ms$ and the lower panels show larger subhalos.  From left to right
subhalos lie closer to the host center, i.e. the background density
increases. In two cases (middle panels of the top and bottom rows) we
also show the best fits obtained without the constraint that the
subhalo mass must be larger than the background within $\rvcmax$
(\textit{long-dashed}). Velocities and radii are given in physical units.}
\label{fig:vcpros}
\end{figure}

The algorithm is based on the friends-of-friends algorithm (FOF)
\citep{Davis1985}, in which particles are linked together if they are
separated by less than some linking length $b \times dx$, where $b <
1.0$ and $dx$ is the mean particle separation.
We simply extend the standard FOF algorithm to include also a
proximity condition in velocity space. Two particles are linked if
\be
(\vec{x_1} - \vec{x_2})^2 / (b \times dx)^2 +  (\vec{v_1} - \vec{v_2})^2 / b_v^2 < 1  \;\; ,
\ee
where $b$ is the standard FOF linking length in units of the mean
particle separation $dx$ and $b_v^2$ is the velocity space linking
length. For $b_v^2 \to \infty$ one recovers the usual real-space
FOF. After the linking stage all groups with at least $N_{\rm min}$
are kept and circular velocity profiles are calculated around the
center of mass of each group using 100 linear spherical bins. Since
6DFOF finds only the peaks in phase-space density we set the radius of
the outermost radial bin to 5 times the half mass radius of the linked
group. The steps described above were implemented in parallel within
PKDGRAV and they run at with vanishing computational expense compared
to the gravity calculations. We omit the computationally expensive
step of removing unbound particles and hence we don't actually
determine subhalo masses.  Instead we estimate $\vcmax$ as a proxy for
mass by fitting the circular velocity profiles using the circular
velocity corresponding to an NFW subhalo embedded in a constant
density background. The three free parameters in these fits are
$\vcmax$ and $\rvcmax$ of the subhalo and the background density. We
minimize the absolute velocity differences. 

Additionally we only allow fits where the subhalo mass enclosed within
$\rvcmax$ is larger than the background mass. This constraint prevents
a few cases (about 2\% of the groups) where small fluctuations in the
background density at large radii are fitted with an extended, diffuse
NFW halo which lies well below the background density at most
radii. Figure \ref{fig:vcpros} shows some example circular velocity
profiles around 6DFOF-structures found within the virial radius of the
$z=86$ SUSY halo. Typical z=0 subhalos are much denser than the local
background and show a clear peak in the total circular velocity
profile. Early subhalos, on the other hand, often lack a clearly
distinguished peak and appear as excess circular velocity above the
linear circular velocity from a constant density background (Figure
\ref{fig:vcpros}).

After some experimentation we found that with $b=0.04$,
$b_v=295\,\kms$ and $N_{\rm min}=32$ the SKID subhalo velocity
function and radial distribution for the $z=0$ cluster B9 are almost
exactly reproduced. They differ significantly only below about $0.025
\vcmaxhost$ which lies below the completeness limit of about $0.05
\vcmaxhost$ in such a simulation
\citep[e.g.][]{Ghigna2000,Diemand2004sub,Reed2005}.  The 6DFOF results
are not sensitive to small (15\%) changes in both $b$ and $b_v$. The
adopted velocity linking of $b_v=295\,\kms$ is about four times
smaller than the mean 1D velocity dispersion within the virial
radius. This suppresses the random links between background particles
even in dense regions were each sphere of radius $0.04\times dx$
contains many particles. Only a prolate structure enclosed in the
inner 15 kpc (about 1.5 percent of the virial radius) is linked
together. In this paper we don't count this central structure as a
subhalo\footnote{SKID produces a similar central object which is
always discarded in this work as well.}.

\begin{figure}[ht]
\includegraphics[width=3.5in]{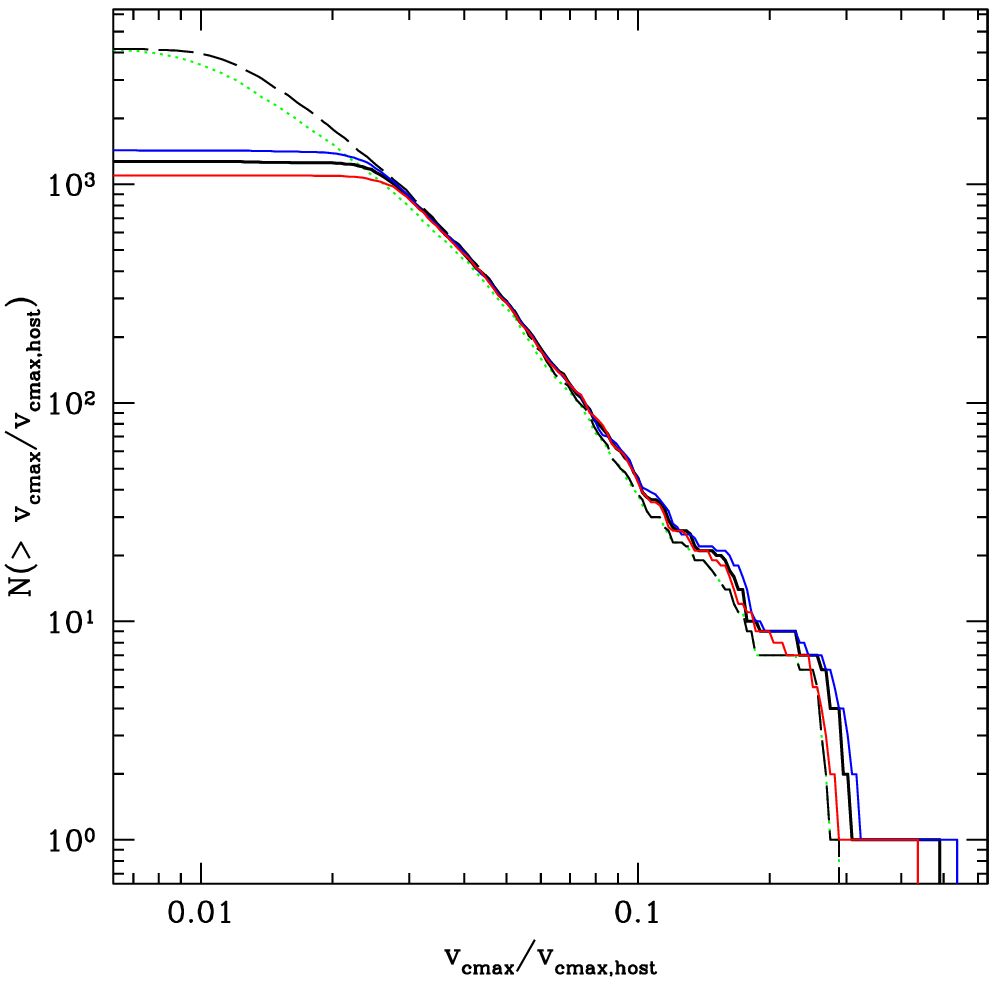}
\includegraphics[width=3.5in]{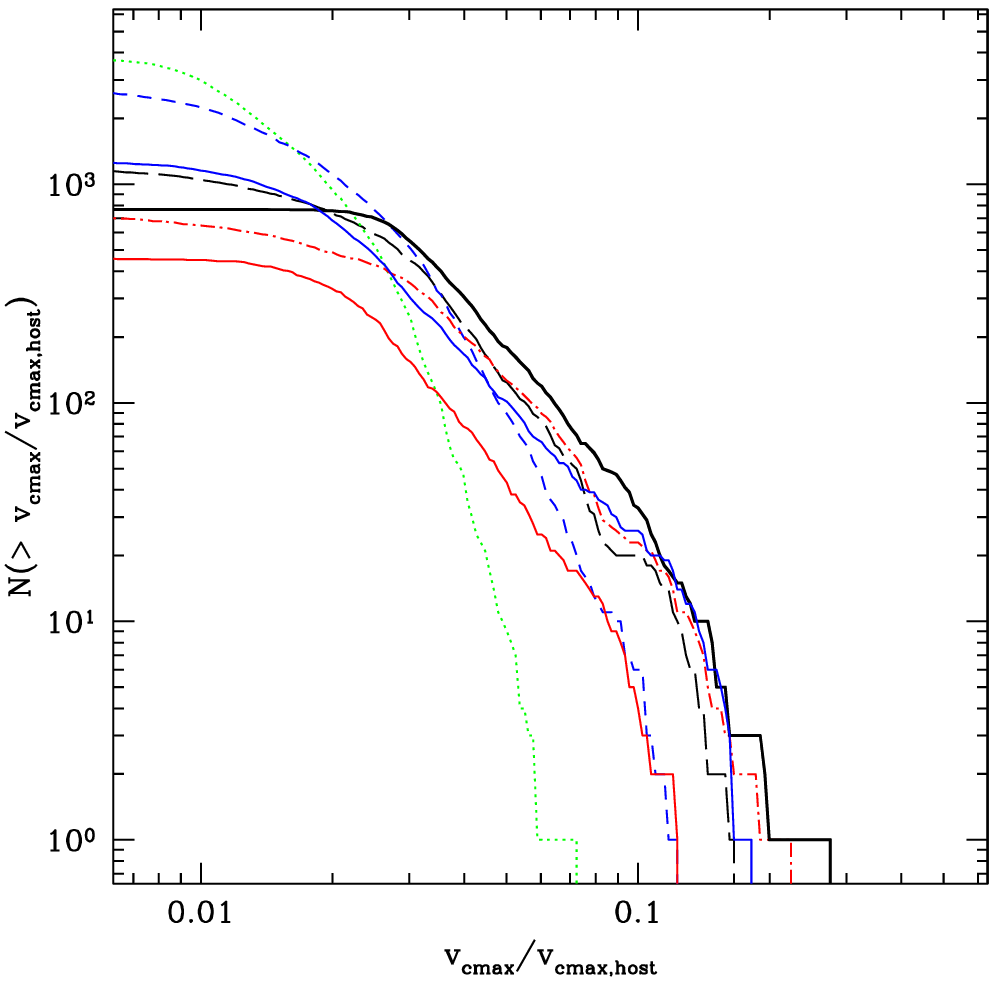}
\caption{SKID and 6DFOF cumulative subhalo velocity functions for the $z=0$
galaxy cluster (\textit{left}) and the $z=86$ SUSY CDM halo
(\textit{right}).  The SKID runs used $N_{\rm min}=10$ and different
numbers of neighbors $s$ for the SPH density calculation: $s=64$
(\textit{dotted}), $s=256$ (\textit{short-dashed}, SUSY halo only),
$s=1024$ (\textit{long-dashed}), and $s=2048$ (\textit{dashed-dotted},
SUSY halo only).  The 6DFOF phase-space group finder was run with
$b=0.04$, $b_v=295\,\kms$ and $N_{\rm min}=32$ (\textit{thick solid})
and 15\% larger (upper \textit{solid}) and 15\% smaller (\textit{lower
  solid}) linking lengths $b=0.04$ and $b_v$. Varying both linking
lengths by 15\% corresponds to a factor of difference 1.5 in both real
and velocity space minimum densities and to a factor of 2.3 in
phase-space density threshold.}
\label{fig:pvelfdep}
\end{figure}

In the SUSY halo at $z=86$ we chose a somewhat larger space linking
length due to the smaller peak overdensities $b=0.057$ and rescale the
velocity linking lengths to $b_v= 4.15\,\ms$ (four times smaller than
the mean 1D velocity dispersion within the virial radius) and we keep
$N_{\rm min}=32$.  Unfortunately just as with SKID also the 6DFOF
subhalo lists at $z=86$ depend strongly on the choice of group-finding
parameters used. A 15 percent increase (or decrease) in both $b$ and
$b_v$ yields a very different velocity function, whereas at $z=0$ our
6DFOF results are unaffected by the same changes in $b$ and $b_v$
(Figure \ref{fig:pvelfdep}). Reducing the 6DFOF linking lengths by 15
percent (a factor of 2.3 in phase-space density threshold) results in
40 percent less subhalos (459 vs. 779) within the virial radius and
the resulting subhalo velocity function is about a factor of five
lower.  Increasing the linking lengths by 15 percent results in an
increase of 68 percent in the number of subhalos (1312 vs. 779), but
many of them have maximum circular velocities well below $0.3\,\ms$,
and show no clear subhalo signal in their total circular velocity
profile $v_{c} (r) = \sqrt{v_{\rm c,sub}^2 + v_{\rm c,BG}^2 }$. Visual
inspection of these small $\vcmaxsub$ structures also shows no
convincing peak in either real or phase-space density. Only for the
highest $\vcmaxsub$ structures do the longer linking length results
agree with those using our fiducial values of $b=0.057$ and $b_v=
4.15\,\ms$.  This strong parameter dependence of 6DFOF results at
$z=86$ is again a consequence of the similar formation times of small
mass, high redshift subhalos and hosts which reduces the contrast in
phase-space density.

When tracing structures over time we set $b\times dx$ and $b_v$ to
constant values in proper coordinates. This corresponds to using the
same proper minimum phase-space density to define structures at
different epochs. This choice is motivated by the fact that the high
(phase-space) density peaks we aim to identify have long decoupled
from the Hubble expansion, i.e. they correspond to the inner cores of
virialized systems. Employing a constant comoving linking length would
result in a decreasing phase-space density threshold. The values we
use are $b\times dx = 0.04 \times 0.111\,{\rm kpc} \times 1/(1+0) =
4.444$ proper kpc, $b_v=295\,\kms$ for the B9 cluster and $b \times dx
= 0.057 \times 0.3\,{\rm pc} \times 1/(1+86.3) = 0.197$ proper Mpc and
$b_v= 4.15\,\ms$ for the SUSY halo.

\end{document}